# Pre-RTL DNN Hardware Evaluator With Fused Layer Support

[1,2] Chih-Chyau Yang and [1] Tian-Sheuan Chang
[1] Institute of Electronics, National Yang Ming Chiao Tung University, Hsinchu, Taiwan
[2] Taiwan Semiconductor Research Institute, NARLabs, Hsinchu, Taiwan
ccyang@narlabs.org.tw, tschang@mail.nctu.edu.tw

*Abstract*—With the popularity of the deep neural network (DNN), hardware accelerators are demanded for real time execution. However, lengthy design process and fast evolving DNN models make hardware evaluation hard to meet the time to market need. This paper proposes a pre-RTL DNN hardware evaluator that supports conventional layer-by-layer processing as well as the fused layer processing for low external bandwidth requirement. The evaluator supports two state-of-the-art accelerator architectures and finds the best hardware and layer fusion group. The experimental results show the layer fusion scheme can achieve 55.6% memory bandwidth reduction, 36.7% latency improvement and 49.2% energy reduction compared with layer-by-layer operation.

## I. Introduction

Deep neural networks (DNNs) have been widely used in modern AI systems and have achieved great success in many applications such as image classification and speech recognition in recent years. However, this high performance comes at the cost of high computation, memory bandwidth and energy consumption, which demands hardware accelerators for real time applications.

Many deep learning accelerators (DLAs) [1]-[3] have been proposed to meet high performance needs with large amount of processing elements and different dataflows. To name a few, Hsiao *et al*. [2] proposes a reconfigurable deep learning accelerator which supports various types of operations. Chang *et al*. [3] proposes a hardware efficient vector-wise convolution accelerator that adopts a 3 × 3 filter optimized systolic array using 1-D broadcast data flow to generate partial sum. This simple and regular data flow results in low area cost and attains high hardware utilization. However, with fast evolving model structures, lengthy hardware design process makes it hard to evaluate the most suitable design in a short time for different models.

To bridge the gap between DLAs and neural network model in the design phase, Cao *et al*. [6] proposes a pre-RTL evaluation framework for neural networks to obtain an optimal configuration of hardware architecture according to the constraints of hardware and performance. Xu *et al*. [7] proposes a DNN chip generator to generate optimized FPGA-based and ASIC-based accelerators and provide performance estimations. However, all these works only consider the conventional layer-by-layer model operation that stores each layer output to the external DRAM and loads it back from DRAM for the current layer input and thus introduces high memory bandwidth requirement. On the other hand, the layer-fusion operation [4][5] fuses multiple layer executions without external intermediate data access, and only requires loading the data from DRAM for the first layer input and storing the data to DRAM for the last layer output in one fusion group. Such operation has become popular to reduce external bandwidth but needs detailed analysis of layer grouping, hardware buffer and computation.

To extend application scope of design analysis, this paper proposes a pre-RTL DNN hardware evaluator that supports conventional layer-by-layer processing as well as the fused layer processing. Besides, this evaluator supports two state-of-the-art accelerator architectures [2][3] instead of the generic non-optimized design as in [6]. Beyond finding the optimal architecture, this evaluator also finds the optimal layer group configuration based on the constraints of latency, area, energy, and memory bandwidth. The evaluation results on VGG-16 show its effectiveness on improvement analysis of bandwidth, latency and energy.

The rest of the paper is organized in the following. Section II introduces the proposed DLA architecture, evaluation metrics, and optimization flow. Section III shows the experimental results. Finally, section IV concludes this paper.

## II. THE PROPOSED ARCHITECTURE, METRICS AND OPTIMIZATION FLOW

### A. The proposed DLA Architecture

Fig. 1 shows the proposed DLA architecture used to estimate the optimal hardware configuration according to the neural network model and user constraints. This DLA consists of the input SRAM, weight SRAM, output SRAM, PE blocks, a system controller and a functional unit for ReLU, batch normalization and pooling operations.

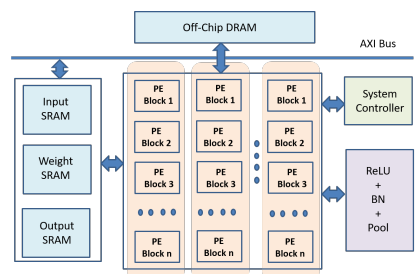

Fig. 1 The proposed DLA architecture

Our proposed DLA architecture can support to implement two state-of-the-art accelerators [2][3] that have optimized to support different layers types with optimized data reuse

C. -C. Yang and T. -S. Chang, "Pre-RTL DNN Hardware Evaluator With Fused Layer Support," 2021 18th International SoC Design Conference (ISOCC), 2021, pp. 83-84, doi: 10.1109/ISOCC53507.2021.9614027.

policies. To implement the design in [2], each PE block contains $F_2$ (rows) by $F_3$ (columns) of PE arrays, and the array of PE blocks can also support the operations for $F_4$ input channels and $F_1$ output channels concurrently. Each PE contains 9 multipliers and an adder tree. To implement design in [3], each PE block becomes $F_2$ by 3 of PE arrays. The array of PE blocks can also support the operations for $F_4$ input channels and $F_1$ output channels concurrently. Each PE contains a multiplier and adder.

*B. Evaluation Metrics*

In this sub-section, the evaluation metrics for memory bandwidth, latency, energy and area for layer fusion model are illustrated below.

For a layer fusion model, assume there are $m$ groups in the model and each group has $n_p$ layers. Each layer consists of convolution operations with $N*N_{ih}*N_{iw}$ input frames, $N*N_{kh}*N_{kw}*M$ filter kernels, and $M*N_{oh}*N_{ow}$ output frames, where $N$ and $M$ indicate the number of input channels and output channels, respectively. The memory bandwidth metric for the layer-fusion operation in a network is listed in Eq. (1).

$$BW = \sum_{p=0}^{m-1}\{\sum_{q=0}^{n_p-1}\{NN_{kh}N_{kw}M\}_{L_{pq}} + NN_{ih}N_{iw} + N_{oh}N_{ow}M\}_{L_p} \quad (1)$$

The latency metric for the layer-fusion operation in a neural network can be formulated as Eq. (2), where $t_{rd\_w}$, $t_{PB}$, $t_{PL}$ and $t_{rd\_IF}$ and $t_{wr\_OF}$ indicate the weight read cycle, input frame read cycle, processing element cycle, pipeline latency cycle, and write output frame cycle, respectively. Eq. (2) is derived with $m$ groups in the model, and $n_p$ layers in each group.

$$L = \sum_{p=0}^{m-1}\{\sum_{q=0}^{n_p-1}\{t_{rd\_W}+t_{PB}+t_{PL}\}_{L_{pq}} + t_{rd\_IF} + t_{wr\_OF}\}_{L_p} \quad (2)$$

The energy metric for the layer-fusion operation, including the energy of external DRAM access, internal SRAM access and processing element operations, is shown in Eq. (3). Each kind of energy is formulated as the access count ($C$) multiplies the energy per memory access or per PE processing ($E$).

$$E = E_{OFCM}C_{OFCM} + E_{SRAM}C_{SRAM} + E_{PE}C_{PE} \quad (3)$$

The area metric for the layer-fusion operation can be formulated as Eq. (4), which contains the area of PE blocks ($A_{PB}$), input frame SRAM ($A_{IFM}$), weight SRAM ($A_{WB}$), and output frame SRAM ($A_{OFM}$).

$$A = A_{PB} + A_{IFM} + A_{WB} + A_{OFM} \quad (4)$$

*C. Optimization Flow*

An optimization flow for layer fusion operation is proposed to obtain an optimal hardware and layer group configuration. We explain this flow below briefly. For each layer group and hardware configuration, the memory bandwidth, area, energy and latency are estimated to check whether they meet the constraints. If the constraints cannot be met, the flow chooses the next configuration for evaluation. If the constraints are met, the configuration index and the corresponding energy are recorded. The hardware and layer group configuration for the smallest energy will be obtained at the end of flow.

III. EXPERIMENTAL RESULTS

VGG-16 is used as an example to demonstrate the proposed evaluator with layer fusion operation. In the following experimental results, VGG-16 is divided into multiple layer fusion groups and each fusion group is separated by the pooling layer. When the constraints of memory bandwidth, latency, energy and area are set to 20 $M$ bytes, 12 $M$ cycles, 65 $mJ$, and 45,000,000 $um^2$, respectively, the optimal hardware configuration $(F_1, F_2, F_3, F_4) = (4, 4, 4, 4)$ can be chosen from predefined configuration set to meet all constraints and obtain the smallest energy. When compared with the layer-by-layer operations, the layer fusion operation achieves 55.6%, 36.7% and 49.2% reduction of memory bandwidth, latency, and energy, respectively. Fig. 2 shows the energy comparisons between fusion-based and layer-by-layer operations. Note that TSMC 40$nm$ fabrication process is used to estimate the DLA area, and the energy per access for $E_{OFCM}$, $E_{SRAM}$ and $E_{PB}$ are set as 1$nJ$, 0.1$nJ$, 0.01$nJ$, respectively.

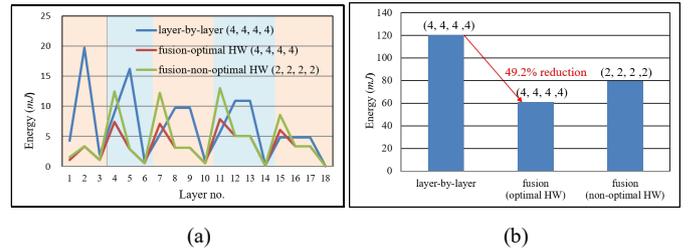

(a) (b)

Fig. 2 Energy comparisons between fusion and layer-based operations

IV. CONCLUSIONS

This paper presents a pre-RTL hardware evaluator to efficiently bridge the gap of DLA and layer fusion model during the design phase. Our proposed evaluator is capable of evaluating and getting the optimal hardware and layer group configuration according to the constraints of latency, area, and energy, memory bandwidth. VGG-16 is used to show the proposed evaluator can function correctly. The experimental results show the layer fusion scheme can achieve the better results in the aspects of memory bandwidth, latency and energy.

REFERENCE

[1] L. Deng, *et al*., "Model compression and hardware acceleration for neural networks: A comprehensive survey," Proceedings of the IEEE, vol. 108, no. 4, pp. 485-532, 2020.

[2] S. Hsiao, *et al*., "Design of a Sparsity-Aware Reconfigurable Deep Learning Accelerator Supporting Various Types of Operations," IEEE Journal on Emerging and Selected Topics in Circuits and Systems, vol. 10, no. 3, pp. 376-387, Sept. 2020

[3] K. Chang and T. Chang, "VWA: Hardware Efficient Vectorwise Accelerator for Convolutional Neural Network," IEEE Transactions on Circuits and Systems I, vol. 67, no. 1, pp. 145-154, Jan. 2020

[4] M. Alwani *et al*., "Fused-layer CNN accelerators," 49th Annual IEEE/ACM International Symposium on Microarchitecture, 2016

[5] C. Lin *et al*., "7.1 A 3.4-to-13.3TOPS/W 3.6TOPS Dual-Core Deep-Learning Accelerator for Versatile AI Applications in 7nm 5G Smartphone SoC," IEEE International Solid- State Circuits Conference, 2020

[6] S. Cao *et al*., "SimuNN: A Pre-RTL Inference, Simulation and Evaluation Framework for Neural Networks," IEEE Journal on Emerging and Selected Topics in Circuits and Systems, vol. 10, no. 2, pp. 217-230, June 2020

[7] P. Xu *et al*., "AutoDNNchip: An Automated DNN Chip Predictor and Builder for Both FPGAs and ASICs," ACM/SIGDA International Symposium on Field-Programmable Gate Arrays, 2020